\documentclass[11pt, a4paper]{article}
\usepackage[utf8]{inputenc}
\usepackage{graphicx}
\usepackage{amsmath}
\usepackage{mathtools}
\usepackage{amssymb}
\usepackage{fullpage}
\usepackage{array, booktabs}
\usepackage{comment}
\usepackage{jinstpub}
\usepackage[sort&compress,numbers]{natbib}
\usepackage{lineno}
\usepackage{dcolumn}

\newcommand{\cf}{C$_3$F$_8$}
\newcommand{\degc}{$^\circ$C}
\newcommand{\cs}{$^{137}$Cs}
\newcommand{\eu}{$^{152}$Eu}
\newcommand{\qseitz}{$Q_{Seitz}$}

\begin{document}
\title{A direct test of Auger cascade induced nucleation from heavy element contamination in \cf\ bubble chambers}
\author[1]{M.~Bressler,}
\emailAdd{mjb536@drexel.edu}
\note[1]{Corresponding Author}
\author{N.~Lamb,}
\author{R.~Neilson}
\author[2]{and S.~Windle}
\note[2]{Now at University of Maryland Dept. of Atmospheric and Oceanic Science}
\affiliation{Department of Physics, Drexel University\\ Philadelphia, Pennsylvania, USA}
\date{\today}

\abstract{
Understanding and quantifying the gamma-induced bubble nucleation background in clean nuclear recoil detection bubble chambers is of utmost importance to bubble chamber based dark matter searches. We present data confirming the hypothesis that large Auger cascades from high-Z elements such as iodine and xenon dramatically increase the response of \cf\ bubble chambers to gamma rays. These tests, performed with a small calibration bubble chamber filled with \cf+$\mathcal{O}$(10) ppm xenon, show that the probability of bubble nucleation scales with the rate of xenon inner-shell vacancies, reaching values >$10\%$ per K-Shell vacancy for Seitz thresholds of interest to future dark matter searches in bubble chambers. We also place an upper limit on bubble nucleation probability for argon Auger events, relevant to large future bubble chambers which may contain some residual atmospheric argon after the active fluid fill.
}

\maketitle
\flushbottom

\section{Introduction}
Bubble chambers have maintained relevance as particle detectors for direct detection dark matter searches \cite{COUPPNIM,60complete,pico60c3f8}, wherein experiments desire to detect rare localized depositions of energy by recoiling nuclei after elastic scatters with dark matter particles. Though bubble chambers were originally used to detect ionizing radiation in high energy particle physics experiments \cite{glaser, bugg, bradner}, experiments using this technology now require detectors which are insensitive to electron recoil (ER) events while maintaining sensitivity to nuclear recoil (NR) events. Recently, PICO performed a reanalysis of various calibration data taken throughout its history in bubble chambers operated with \cf\ to empirically model the response of such detectors to gamma ER events  \cite{ERpaper}. The study found two main trends: a model for ``pure \cf'' which included the definition of a new kind of thermodynamic threshold, and a separate model for \cf\ contaminated with trace amounts of iodine, leftover in the detectors from previous experiments using CF$_3$I. In pure \cf, the amount of energy deposited determines the probability of bubble nucleation by ionization, but in iodine-contaminated chambers, along with some qualitative results in a bubble chamber with tungsten contamination \cite{baxterthesis}, the model indicates that Auger cascades initiated by inner-shell photoabsorptions on the high-Z contamination are to blame for increased ER sensitivity. This conclusion also explains why \cf\ has overall better ER rejection than CF$_3$I. The pure \cf\ model's predictive power was tested directly with two small test chambers: Gunter (at the University of Chicago) and the DBC (\emph{D}rexel \emph{B}ubble \emph{C}hamber, at Drexel University). Both of these calibrations agreed with the historical calibration data taken in previous test chambers and dark matter detectors. 

Before now, no test was performed to independently reproduce the model in \cf\ with high-Z contamination, and the level of iodine in the contaminated chambers could not be accurately determined, leaving large uncertainties on the fit parameters. We have now performed such a test with the DBC, using xenon ($Z=54$) as a stand-in for iodine ($Z=53$), added to the \cf\ in known small amounts.\footnote{As an interesting historical note, this is not too dissimilar from Pierre Auger's 1920s cloud chamber experiments, where small amounts of photoabsorbing elements, such as noble gasses, were diluted in the cloud chamber with hydrogen, which does not show much photoelectric absorption. Thus localized Auger events, including multiple-ionization (the ``double-photoelectric effect'') could be studied, showing both the initial photoelectron track, and the tracks of the multiple Auger electrons \cite{augercloudchamber}.} Xenon and iodine have a similar electronic structure, thus they produce similar Auger cascades of about ten electrons from inner-shell (K and L) photoabsorption events \cite{XenonAndIodineAugerSpectra}, at roughly the same rate; xenon's K-Shell binding energy is 34.6~keV while iodine's is 33.2~keV \cite{bindingenergies}, both well above the necessary energy deposition for bubble nucleation ($<5$~keV in our experiments). We also present a test of the effect of Auger cascades from argon K-shell vacancies (binding energy 3.2~keV \cite{bindingenergies}) on bubble nucleation. We expect that unlike iodine, xenon and argon will be easy to remove after these experiments by emptying the chamber and pumping it to vacuum, since xenon is less reactive and thus less likely to stick to the surfaces.

The operational pressure and temperature of a bubble chamber set its thresholds for particle detection. The threshold used for nuclear recoil events is referred to as the Seitz threshold \qseitz, after Frederick Seitz, who described the operational theory of bubble chambers in 1958 \cite{Seitz}. The formula used for the Seitz threshold in the current work is that derived by COUPP, which is updated from Seitz' original work to more accurately represent the energy used to create a bubble \cite{COUPPCF3I, ERpaper}. In pure \cf, the threshold for nucleation by gammas is not well-predicted by \qseitz, and is instead given by a threshold referred to as $E_{ion}$ \cite{ERpaper}, and the probability of a gamma interaction creating a bubble is dependent on the amount of energy deposited in the interaction (as opposed to the NR case, where to a good approximation the bubble chamber acts as a threshold detector, with 100\% efficiency well above \qseitz\ and 0\% efficiency below). Because \qseitz\ and $E_{ion}$ are not identical, we can decouple a bubble chamber's sensitivity to NRs (from, e.g., neutrons or WIMPs) and its sensitivity to ERs (i.e. gammas). However, in the case of iodine-contaminated bubble chambers, and those operated with CF$_3$I, the sensitivity to gammas appears to trend with \qseitz\ rather than $E_{ion}$, an effect which is attributed to the NR-like effective stopping power of Auger cascades occurring after inner-shell photoabsorptions on the heavy atoms. Auger cascades have been proposed as a mechanism contributing to bubble nucleation even in the era of bubble chambers as tracking detectors, since each low-energy electron emitted in the cascade contributes to local heating \cite{tenner, simultaneouscharge}, and the dense Auger cascade energy depositions mimicking nuclear recoil signals are also a known potential background to xenon time projection chamber (TPC) dark matter searches \cite{xelda}.

\section{Experimental Method}
\begin{figure*}[h!]
    \centering
    \includegraphics[width=\textwidth]{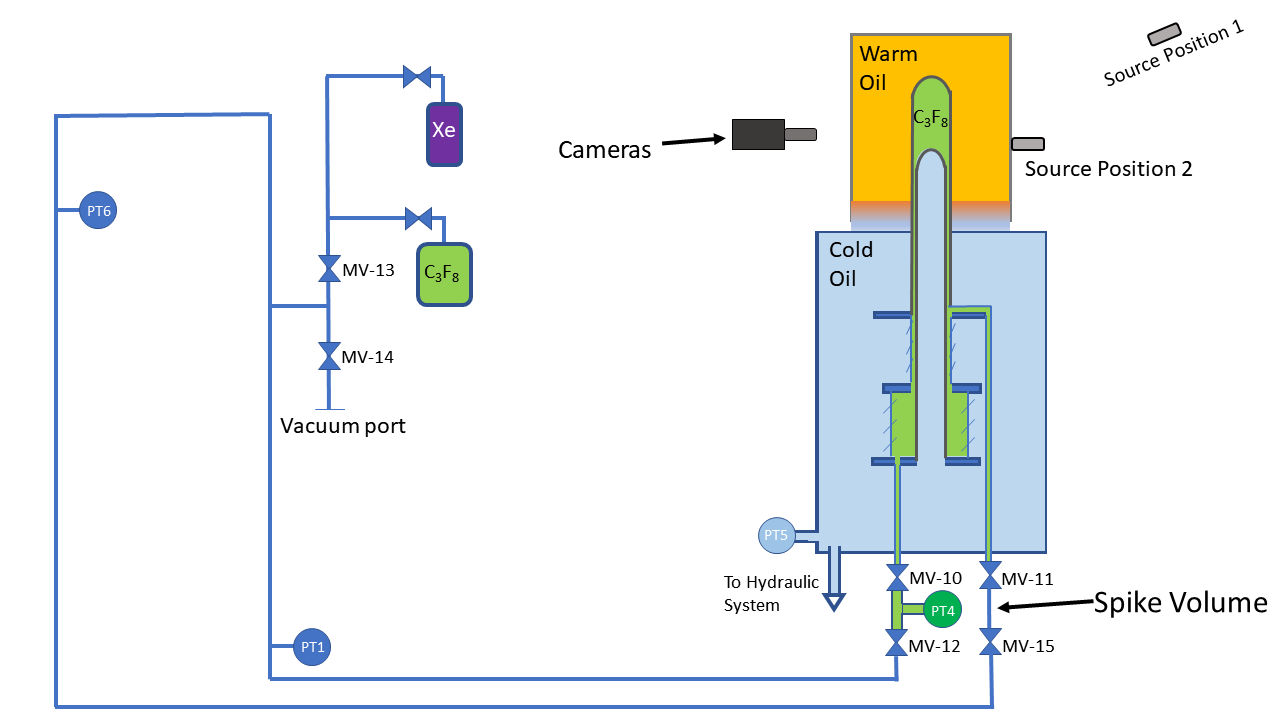}
    \caption{A simple schematic of the DBC, showing the gas fill system and source positions used for the data presented here. In position 1, the source is $27.7\pm 0.5$~cm from the center of the chamber, in position 2 it is $6.7\pm0.5$~cm.}
    \label{fig:schematic}
\end{figure*}

The DBC is a calibration/prototype detector operated above ground at Drexel, and was described in detail in ref. \cite{DBCpaper}. It contains an active volume of $\sim$40~mL of \cf, which has been kept free of high-Z element contamination until the tests presented here. For these tests, a second fill line was added to the \cf\ fill system to allow us to add compressed gases to the chamber; a simplified schematic of the gas fill system and chamber layout is shown in figure \ref{fig:schematic}. The ``spike volume'' is a straight section of tube between manual valves MV-11 and MV-15, designed to trap a specific quantity of gas which is subsequently dissolved into the \cf. 

Additional changes to the apparatus after the work presented in ref. \cite{DBCpaper} include a new active fluid pressure transducer, which is immersed in \cf\ inside a mini-ConFlat nipple, and updated control parameters to increase the fraction of time the detector is live. For this work, compressions between events are 15~seconds long, with a 60~second compression every ten events, the maximum time the chamber is allowed to be expanded is 600~seconds, and the compressed pressure has been reduced to 175~psia.\footnote{Previously, compressions were at 200~psia, each compression was 30~sec long, a long compression of 300~sec was performed every 30 events, and the maximum expanded time was 300~sec. We performed tests to determine whether such long compressions (i.e. dead time) and high compression pressure were necessary for the stability of the chamber, and found that they could be reduced so that we have less required detector dead time per event and put less stress on the equipment. The old parameters were motivated by previous COUPP and PICO bubble chambers which used a water buffer fluid, and longer compressions were necessary to ensure that active fluid that boiled into the water had time to condense and drop back into the bulk region.} The time during the compressed period, the time that it takes for the pressure to reduce to and stabilize at the expansion set point, and the time that it takes to recompress after a trigger is considered dead. Thus detector ``live'' time is counted only in the fully expanded low-pressure state, and we refer to the total time that the detector is operating (cycling) as ``real'' time. The live time fraction is then defined as the live time of a run divided by its real time. For background runs with the control parameters used in this work, the live time fraction is typically $\sim70\%$.

As in ref. \cite{DBCpaper}, two cameras monitor the chamber to provide the primary trigger to save the data and compress the chamber. Photographs of the chamber at the time of the trigger are analyzed to find the number of bubbles in the event and their locations. For all data in the present work, events with more than one bubble are cut from the analysis; multiple-bubble events are very likely neutrons which are not relevant to the Auger cascade experiments. Rates are thus presented as single bubble events per live hour.

The \cf\ pressure transducer PT4 was initially used for feedback to control the pressure in the argon data, but for the xenon data the pressure was controlled with feedback from the hydraulic fluid's pressure transducer PT5. In the argon data, PT5 is used for threshold calculations, but in the xenon data PT4 is used, so the transducers effectively swapped roles between the data sets\footnote{During the argon tests, the PT4 transducer was found to be unreliable due to electrical connector corrosion. PT4's connector was repaired before proceeding to the xenon data, but the decision to regulate on PT5 (the transducer measuring the hydraulic pressure on the opposite side of the pressure-balancing bellows to the \cf) for the xenon data was made, out of necessity to regulate on a reliable physical pressure, to mitigate the possibility of PT4 failing again and the regulation being performed with a non-physical value of the pressure.}. 

During normal operation, MV-15 and MV-12 are closed, containing the \cf\ above them, with MV-10 and MV-11 open. To add xenon or argon ``contaminant'' gas to the chamber, MV-11 is closed and MV-15 is opened, and the fill system is pumped to vacuum ($<$100~mTorr) with a scroll pump. Gas is then added from the cylinder, via a regulator, to a determined pressure (read out by PT1, PT6, and the regulator gauge), and MV-15 is closed, trapping the gas in the spike volume. After trapping the desired amount of gas between MV-15 and MV-11, MV-11 is opened, allowing the \cf\ above to fill the spike volume, and the gas to dissolve into the \cf. To encourage mixing of the added gas, cooling to the spike volume is turned off, allowing the stainless steel tube to rise above 0\degc, and the chamber is expanded and recompressed at least 100 times with the spike volume at this elevated temperature. During these expansions, fluid in the spike volume (and other locations around the cold region) boils, mixing the \cf\ and contaminant gas. The spike volume is then cooled back down, and the temperatures in the chamber are allowed to stabilize for a period of at least a few hours before operation resumes.

The spike volume is $2.6\pm0.3$~mL, and the total volume of \cf\ during normal operation is 263~mL including both the active and inactive regions, so for every absolute psi of xenon in the spike volume during a xenon addition, we add $2.7\pm0.3$~ppm of xenon by mass to the active fluid. These uncertainties stem from an overall 10\% uncertainty on dimensions of components used to determine the volumes in the system.

Electron recoil events are induced with \cs\ and \eu\ gamma sources. The electron capture decay of \eu\ frequently results in emission of 40-46~keV X-rays with high cross-sections for K-shell photoabsorption on xenon, thus more efficiently induces Auger cascades than \cs\ (see Sec. \ref{section:simulations}). Comparing the sources allows a test of the hypothesis that K-shell events are responsible for bubble nucleation, rather than the rate of energy deposition (as in the pure \cf\ case). As an additional test, for some data runs a 0.25~mm thick tin sheet is inserted to selectively absorb these X-rays and other low-energy gammas.

The \cs\ source is the same as that used with the DBC in refs. \cite{DBCpaper, ERpaper}; with label activity 99.11~$\mu$Ci and reference date 1~February, 2011, giving mean activity during this test of 79.84~$\mu$Ci. The \eu\ source has a label activity of 10~$\mu$Ci, but no reference date. Calibration against known sources using a NaI detector at Drexel yields an activity of $1.59\pm0.24$~$\mu$Ci on February 8, 2021, corresponding to a mean activity during our data collection $1.64\pm 0.25$~$\mu$Ci.

The \cs\ source is used in two different positions (shown in figure \ref{fig:schematic}, named position 1 and position 2), while the \eu\ source is only used in position 2, due to its low activity. In position 1, the \cs\ is placed inside a lead collimator aimed at the active region of the bubble chamber; in position 2, the sources are placed without collimation on an acrylic paddle butted against the wall of the warm bath surrounding the active region, and the paddle is marked to ensure reproducible source placement for each run. For some data from early in the test, the \eu\ source was used inside its plastic case, which potentially had a small effect of absorbing gammas, so the case was removed for the majority of the data. Simulations with and without the plastic case indicated that, to within uncertainties, the effect of the plastic was insignificant, so we do not treat those data differently. 

\section{Simulations}
\label{section:simulations}
\begin{figure}
    \centering
    \includegraphics[width=\textwidth]{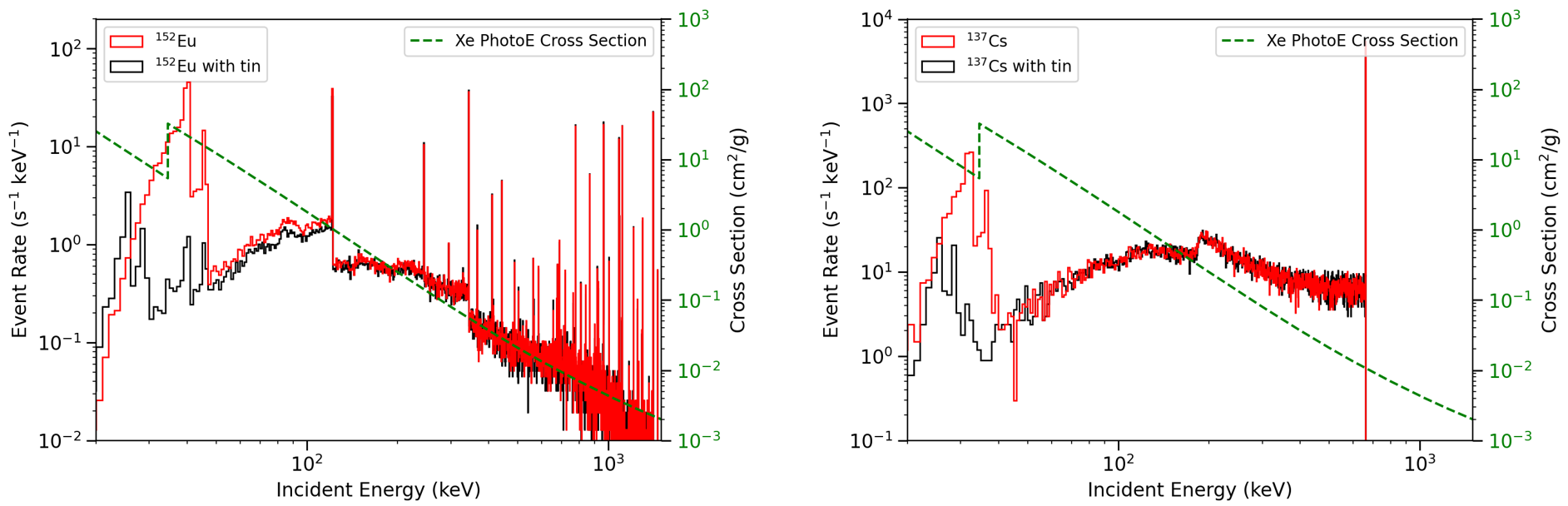}
    \caption{The spectrum of gammas from the \cs\ and \eu\ sources in position 2, with (black) and without (red) the 0.25~mm tin sheet. With both sources, the rate of inner shell vacancies is reduced when the tin is in place while the rate of energy deposition remains relatively constant. The reduction in inner-shell vacancies is more dramatic with \eu\ because a larger portion of the \eu\ low-energy spectrum overlaps with a region of enhancement around the K-shell binding energy of the xenon photoabsorption cross-section (shown in green). The tin also introduces x-rays that can be seen in the spectra as series of peaks around 20~keV.}
    \label{fig:spectra}
\end{figure}

Gamma ray interactions in the bubble chamber were simulated with GEANT4 version 10.03.03 \cite{geant4}, using the \texttt{Shielding.icc} physics list with the default electromagnetic model replaced by the Livermore model, which provides more accuracy for low energy interactions. Figure \ref{fig:spectra} shows the simulated energy spectra of photons that interact within the detector for \cs\ and \eu\ sources with and without tin absorber. The photoelectric cross section of xenon peaks at the xenon K-shell binding energy (34.5 keV), so a lower rate of incident gammas just above that threshold will result in a lower rate of k-shell vacancies, as is seen with the \eu\ simulations with tin when compared to those without tin.

We record both the total energy deposited in the chamber, and the number of inner-shell vacancies in high Z contaminants, which are expected to de-excite via Auger cascades. Electron vacancies are counted from both photoelectric and Compton interactions; photoelectric interactions account for $\sim$99\% of K-shell vacancies in xenon, however Compton scattering contributes significantly to xenon L-shell and argon K-shell events. Inner-shell vacancies due to the photoelectric effect are directly counted in the simulation, and inner-shell vacancies due to Compton scattering are estimated by counting the total number of Compton scatters on contaminant atoms and assuming that all electrons are equally likely to be ionized so long as the photon deposited more energy than the binding energy of the electron. The rate of inner-shell events is found to be linear for small concentrations of xenon ($\lesssim$1000~ppm) and argon ($\lesssim$5000~ppm). Thus, we simulate the chamber with 1000~ppm xenon for the xenon spike analysis and with 5000~ppm argon for the argon spike analysis, and scale the simulations results linearly for each data point. This scheme was also used for the simulations in ref.~\cite{ERpaper}. To within uncertainties, the overall rate of energy deposition in the chamber does not depend on the concentration of xenon or argon.

Table \ref{table:simrates} shows the rates of energy deposition and K- and L-Shell vacancies in the chamber. From these simulation results, we expect the rate of Auger cascades in data without tin to be $5.9\pm1.3$ times the rate with tin for \eu, but the same ratio for the total rate of energy deposition is only $1.1\pm0.2$. In addition, the ratio between inner shell vacancies for \eu\ in Position 2 and \cs\ in Position 1 is $2.5\pm0.3$, while the same ratio of energy deposition for these simulations is $0.40\pm0.05$. These ratios give us the opportunity to test the Auger cascade model independent of source activity.

\begin{table}
\centering
\begin{tabular}{lD{,}{\pm}{-2}D{,}{\pm}{-2}D{,}{\pm}{-2}}
\toprule
Source Configuration & \multicolumn{1}{c}{K-Shell Rate} & \multicolumn{1}{c}{L-Shell Rate} & \multicolumn{1}{c}{Energy Deposition Rate}\\
 & \multicolumn{1}{c}{[h$^{-1}$~ppm$^{-1}$]} & \multicolumn{1}{c}{[h$^{-1}$~ppm$^{-1}$]} & \multicolumn{1}{c}{[MeV~s$^{-1}$]} \\
\hline
\hline
\cs, Position 1 & 22.2 , 1.3 & 11.3 , 0.9  & 204.8 , 8.6\\

\cs+Tin, Position 1 & 17.4 , 1.2 & 4.3 , 0.5 & 189.9 , 8.0\\

\cs, Position 2 & 199.4 , 30.4 & 121.0 , 18.7 & 2130 , 320\\

\cs+Tin, Position 2 & 149.8 , 23.0 & 38.1 , 6.3 & 2106 , 314\\
\hline

\eu, Position 2 & 68.8 , 10.3 & 15.0 , 2.3 & 81.3 , 12.1 \\

\eu+Tin, Position 2 & 11.6 , 1.8 & 2.7 , 0.5 & 74.5 , 11.1\\
\bottomrule
\end{tabular}
\caption{Rates of energy deposition and xenon inner-shell vacancies, from the GEANT4 simulation. The uncertainties on the simulated rates come from combining the statistical uncertainty and a 5~mm uncertainty in distance from the center of the chamber, calculated assuming that the gamma flux through the chamber falls as $1/r^2$. The source positions are illustrated in figure \ref{fig:schematic}.}
\label{table:simrates}
\end{table}

\section{Data Analysis and Results}
Analysis is performed according to the procedures and event selection criteria of ref \cite{DBCpaper}. Cuts include: removal of events in which a bubble rises into the view of the cameras after being nucleated in the temperature transition region, removal of events in which a bubble was nucleated during the period where the chamber was in the process of expanding and had not yet reached the pressure setting, and removal of events in which more than one bubble was nucleated. No attempt is made at acoustically identifying alpha decays. After all cuts, the xenon spike data contains a total of 64000 single bubbles and 1563 live hours, collected over a detector operation time of 2544 real hours. Data was taken at a range of Seitz thresholds, to explore the parameter space of Auger cascade sensitivity. Pressure setpoints ranged from 30 to 50~psia. Most of the data was taken at a temperature of 19\degc\ (79 gamma gamma source runs and 32 background runs, ranging from 0~ppm to 32~ppm xenon), and some at 17\degc\ (18 gamma source runs and 6 background runs, 32~ppm xenon only), resulting in Seitz thresholds between 1.5 and 4.8~keV. For each data point, we calculate the average temperature and pressure during the runs, and those values are used to calculate the thermodynamic parameters, including \qseitz, using the NIST REFPROP database \cite{refprop}. 

\begin{figure}
    \centering
    \includegraphics[width=\textwidth]{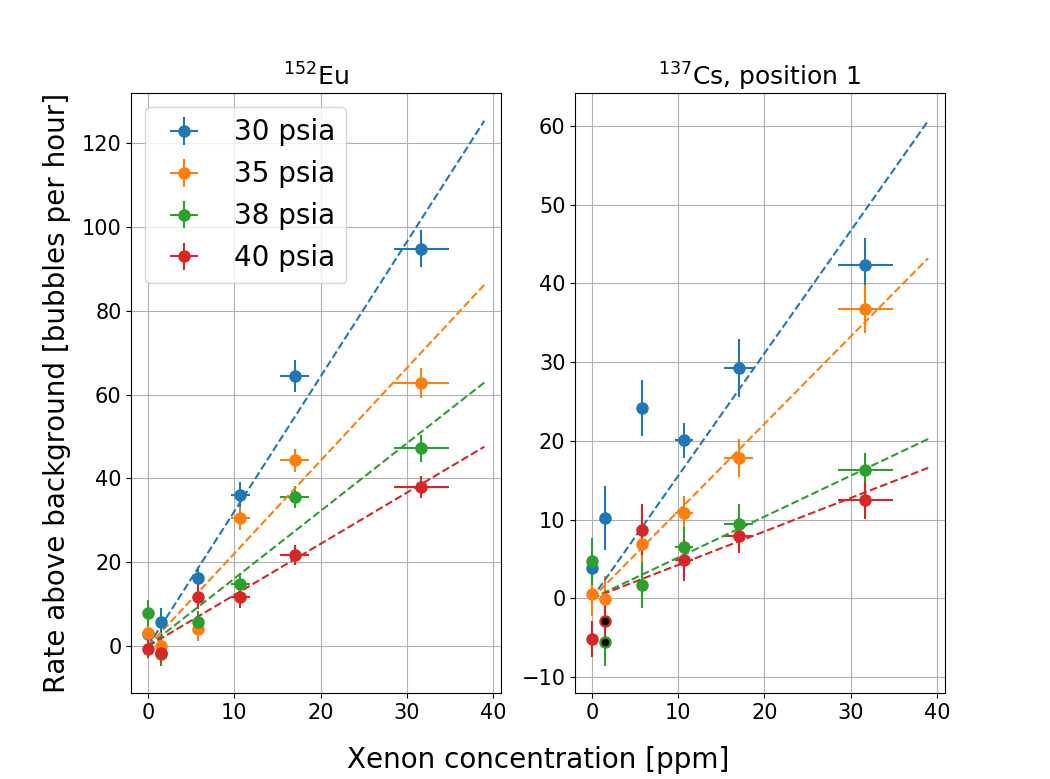}
    \caption{Bubble rates in the DBC for data taken at 19\degc, showing linearly increasing trends with xenon concentration in the \cf; data sets are grouped by expansion pressure setpoint, spanning nominal values of \qseitz\ between 1.56~keV at 30~psia and 2.22~keV at 40~psia. Horizontal error bars show the 10\% uncertainty on the concentration of xenon, but are not used in the linear fits (dashed lines). For \cs, the two points with black centers indicate data taken with xenon but whose entire 1$\sigma$ regions lie below the horizontal axis, so they are not visible in figure~\ref{fig:fullodrPvsf}. Fits are constrained to be zero at 0~ppm. Due to variations in the actual pressure and temperature during the runs, small deviations from true linear trends are expected (i.e. all points of the same color are not at exactly the same Seitz threshold). The linear fits shown here are not used elsewhere in the analysis, and only serve to demonstrate the trend of rate with xenon concentration.}
    \label{fig:ratelinear}
\end{figure}

Background data was taken at each pressure and temperature setpoint, for each concentration of xenon, and the rate was found to be consistent across all xenon concentrations and thresholds, indicating that the chamber's sensitivity to background events does not change with the contamination.\footnote{The background rate is assumed to be made up of mostly cosmogenic neutrons, whose scatters in the chamber deposit energy well above \qseitz, so adjusting \qseitz\ by a few keV does not significantly change the chamber's sensitivity to these background events.} The mean background rate was 28.6~h$^{-1}$. With the gamma sources in place, a linear increase in the bubble rate above background with xenon concentration is visible, shown in figure \ref{fig:ratelinear} with background subtraction. The increase in ER sensitivity is clearly evident at concentrations of xenon on the 10~ppm level and the rates with \cs\ are overall lower than the rates with \eu, consistent with the rates of K-shell vacancies rather than the rate of energy deposition.

\subsection{Comparison to the PICO Analysis}
To further quantify the increase in ER sensitivity, we fit to the model presented in ref. \cite{ERpaper}, where the rate of bubbles per K-Shell vacancy is given as a function of a stopping power labeled $f(P,T)$. For the Auger cascade nucleation model, the energy term in the numerator of the stopping power is \qseitz:
\begin{equation}
    f(P,T) = \frac{Q_{Seitz}}{r_l\rho_l}
\end{equation}
where $\rho_l$ is the density of the liquid, and $r_l$ is the radius of a sphere which contains the superheated liquid that will form the critically-sized bubble when it boils. The critical radius is $r_c = 2\sigma/(P_b-P_l)$, where $\sigma$ is the surface tension, and the denominator is the difference between the vapor pressure in the bubble and the pressure of the superheated liquid, so
\begin{equation}
    r_l = r_c \left(\frac{\rho_b}{\rho_l} \right)^{1/3}
\end{equation}
where $\rho_b$ is the density of gas in the critical bubble. The probability of nucleation per simulated K-shell vacancy (i.e. presumed Auger cascade) is then an exponential with two free parameters:
\begin{equation}
    \mathcal{P} = A e^{-Bf(P,T)}.
\end{equation}
However, it is most useful to show the inverse of the multiplier in the exponent; $B^{-1}$ has units of stopping power and comes out on $\mathcal{O}(100)$~MeV~cm$^2$~g$^{-1}$, and may be related to the effective stopping power of K-shell Auger cascades in liquid \cf. The total stopping power of a 1~keV electron in liquid \cf\ is about 90~MeV~cm$^2$~g$^{-1}$ \cite{NISTESTAR}. 

\begin{figure}
    \centering
    \includegraphics[width=\textwidth]{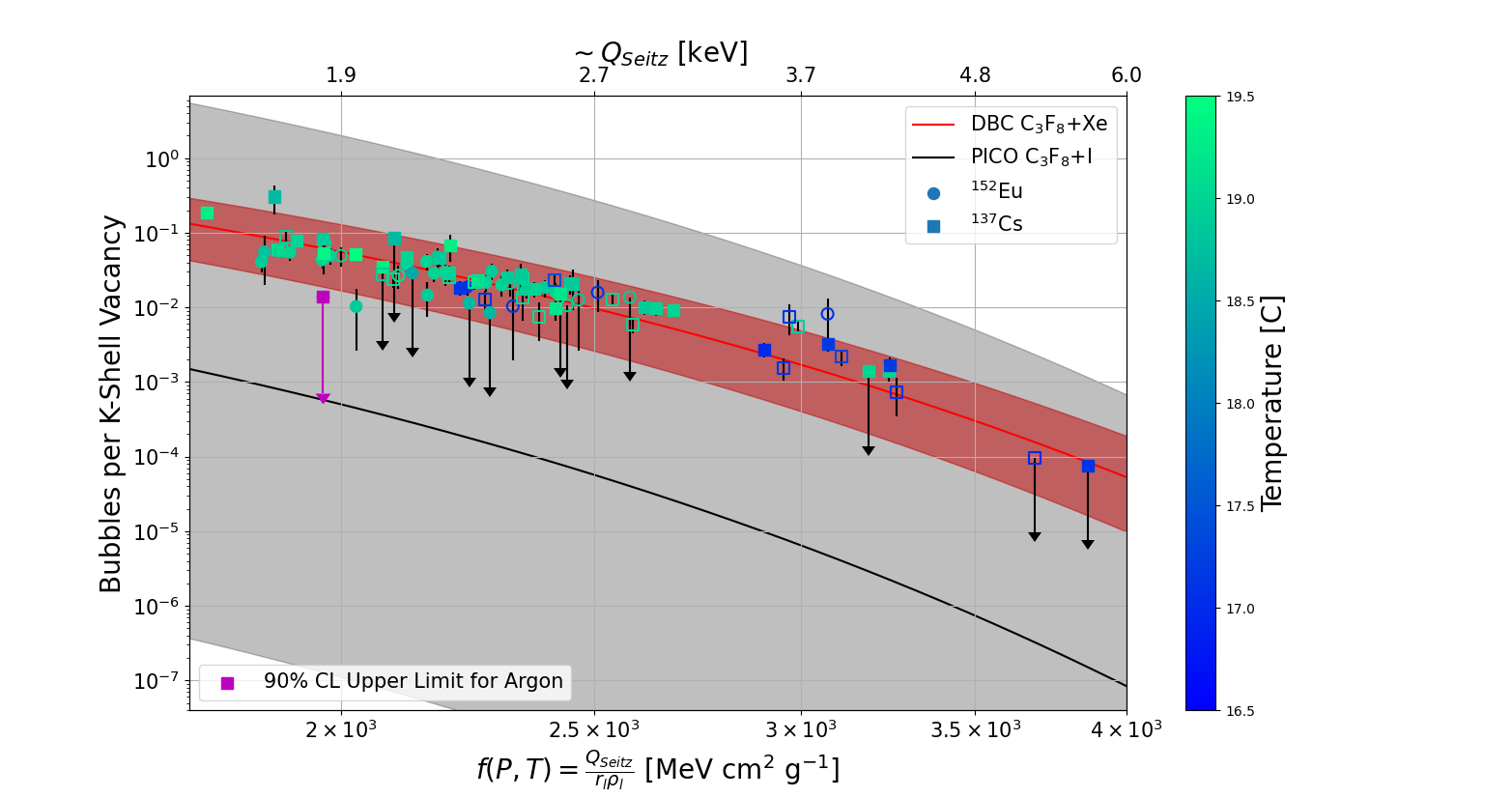}
    \caption{All DBC data with xenon, showing good agreement with the exponential decay model. Data with tin is shown as unfilled markers, filled markers represent data taken without tin. The color shows the mean temperature; data taken at different temperatures but the same $Q_{Seitz}/r_l\rho_l$ agree well. Down-pointing black arrows indicate points consistent with zero at the 1$\sigma$ level, and the markers for these points are placed at the 1$\sigma$ upper limit. Our \cf+Xe data all lie above the PICO \cf+I best fit, however they are well within the 1$\sigma$ region (shaded gray) based on the large error bars on PICO's fit parameters, which result in probabilities spanning more than 6 orders of magnitude. The red shaded region is 1$\sigma$ around the DBC best fit, based on uncertainties on our fit parameters. The magenta point shows the 90\% confidence level upper limit on argon K-shell nucleations set in section \ref{section:argonspike}. Note that the top horizontal axis showing the value of \qseitz\ for these points is approximate; the conversion from \qseitz\ to $\frac{Q_{\text{Seitz}}}{r_l\rho_l}$ depends on the temperature, and we have used 19\degc\ in this case. The approximation is good over the temperature range of our data ($\sim1\%$) but should not be taken too far from 19\degc.}
    \label{fig:fullodrPvsf}
\end{figure}

Figure \ref{fig:fullodrPvsf} shows all of the DBC data with xenon and a fit to the exponential model. The fit is performed in Python with the scipy orthogonal-distance regression (ODR) module \cite{scipy, odr}. The ODR module takes in all uncertainties, combined when they are along the same axis, and uses them to weight the fit and estimate the uncertainties on the best-fit parameters. The uncertainties in the vertical direction include the standard statistical uncertainty in the data, the absolute concentration of xenon (10\%), the activity of each source (\cs: 3\%; \eu: 15\%), statistical uncertainty on the simulated rates (varies by point), and the 5~mm position uncertainty (translated to an uncertainty on the simulated rate of events, varying among the different sources and positions). In the horizontal direction, 10\% uncertainty is assigned to the values of $f(P,T)$ to account for variations and uncertainty in the pressure and temperature during the runs. All uncertainties are treated as uncorrelated. The dominant uncertainties are the uncertainty on $f(P,T)$ and the statistical uncertainty in the data, which are naturally uncorrelated; treating other uncertainties as fully correlated or fully uncorrelated does not significantly change the fit results. Two points taken with \cs\ in position 1 at the lowest xenon concentration, 1.5~ppm, do not appear on the plot in figure \ref{fig:fullodrPvsf} (but are included in the fit) because they lie 1--2$\sigma$ below the horizontal axis. These can be seen in terms of rates in the right panel of figure \ref{fig:ratelinear} as the green and red points with black centers. Table \ref{table:fitparams} compares our best fit parameters to those of ref. \cite{ERpaper}. The uncertainties on our fit parameters are much smaller than those in the PICO fit because we know the xenon concentration in the chamber, whereas PICO was unable to accurately determine the iodine concentration in the contaminated chambers.

\begin{table}
\centering
\begin{tabular}{|c|c|c|}
\hline
~ & $A$~[K-phot$^{-1}$] & $B^{-1}$~[MeV~cm$^2$~g$^{-1}$] \\
\hline
\hline
DBC \cf+Xe & $58\pm 30$ & $287.7\pm 18.8$\\
\hline
Ref. \cite{ERpaper} \cf+I & $3\times10^{0\pm3.3}$ & $230\pm20$\\
\hline
\end{tabular}
\caption{Fit parameters, comparing our \cf+Xe data taken with the DBC to PICO's fit parameters with iodine-contaminated \cf. The value of $A$ found here fits well within the range given by PICO, and the value of $B^{-1}$ is slightly higher than that of ref. \cite{ERpaper}, by $2.1\sigma$. In the PICO data, the absolute iodine concentration was unknown, leading to the large uncertainty in the fit.}
\label{table:fitparams}
\end{table}

\subsection{Source Activity-Independent Analysis}
\begin{figure}
    \centering
    \includegraphics[width=\textwidth]{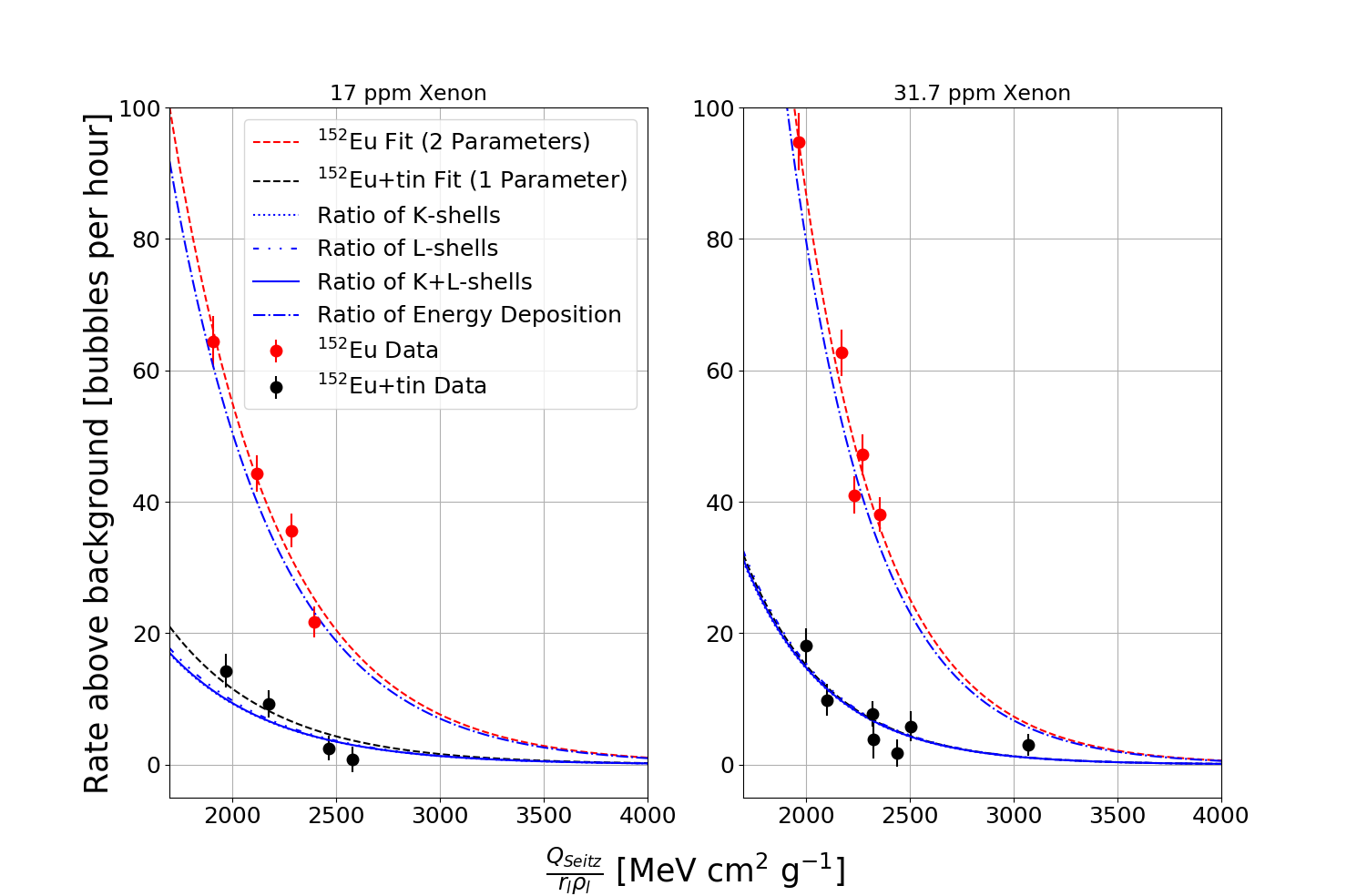}
    \caption{Bubble rates in DBC data with the \eu\ source, with and without tin (black and red points, respectively, as in figure~\ref{fig:spectra}), compared to the expectation for the bubble rate if bubbles are nucleated according to the rate of energy deposition (blue dot-dashed line), or the rate of Auger cascades from K-shell vacancies (blue dotted line), L-shell vacancies (blue dot-dot-dashed line), or both K- and L-shell vacancies (solid blue line). Red dashed lines are exponential fits to each no-tin data set, and the rate expectations (blue lines) are calculated by multiplying the red dashed lines by the simulated ratios, with tin to without tin, of the rates of energy deposition and the rate of Auger cascades. Black dashed lines are exponential fits to the data with tin, allowing only the coefficient to float while constraining the decay constant to be the same as the red-dashed line, so the fit determines the best ratio between the black and red points.}
    \label{fig:ratioseu}
\end{figure}

As a second check, we perform an analysis that does not depend on the source activities, using only the ratios of the rates of events with and without the tin sheet in front of the sources. The bubble rate in data with tin is reduced, consistent with the expected decrease in the rate of Auger cascades (inner-shell vacancies), and inconsistent with the change in the rate of energy deposition. This effect is most pronounced in \eu\ data. Figure \ref{fig:ratioseu} shows bubble rates for 17 and 31.7~ppm \eu\ data with and without tin, along with predictions of the with-tin rate assuming the rate scales with either total energy deposition or inner-shell vacancies. We show the expectations for the with-tin rate based on an exponential fit of the form $y=ae^{bx}$ to the without-tin rate multiplied by the simulated ratio of inner-shell events with tin to those without. The data with tin fit the predicted rates based on the ratio of Auger cascades initiated by K-shell vacancies well, with $\chi^2/NDF = 6.2/4$ for 17~ppm, and $\chi^2/NDF = 7.4/7$ for 31.7~ppm. The hypothesis that the rate scales with the rate of ER energy deposition is a bad fit to the with-tin data; $\chi^2/NDF = 543/4$ for 17~ppm, and $\chi^2/NDF = 1545/7$ for 31.7~ppm. For comparison, the data with tin is fit to an exponential function with only the coefficient as a free parameter while constraining the parameter in the exponent to be the same as that of the best fit to the without-tin data, so the coefficient parameter contains the best-fit ratio for our data. 

\begin{figure}
    \centering
    \includegraphics[width=\textwidth]{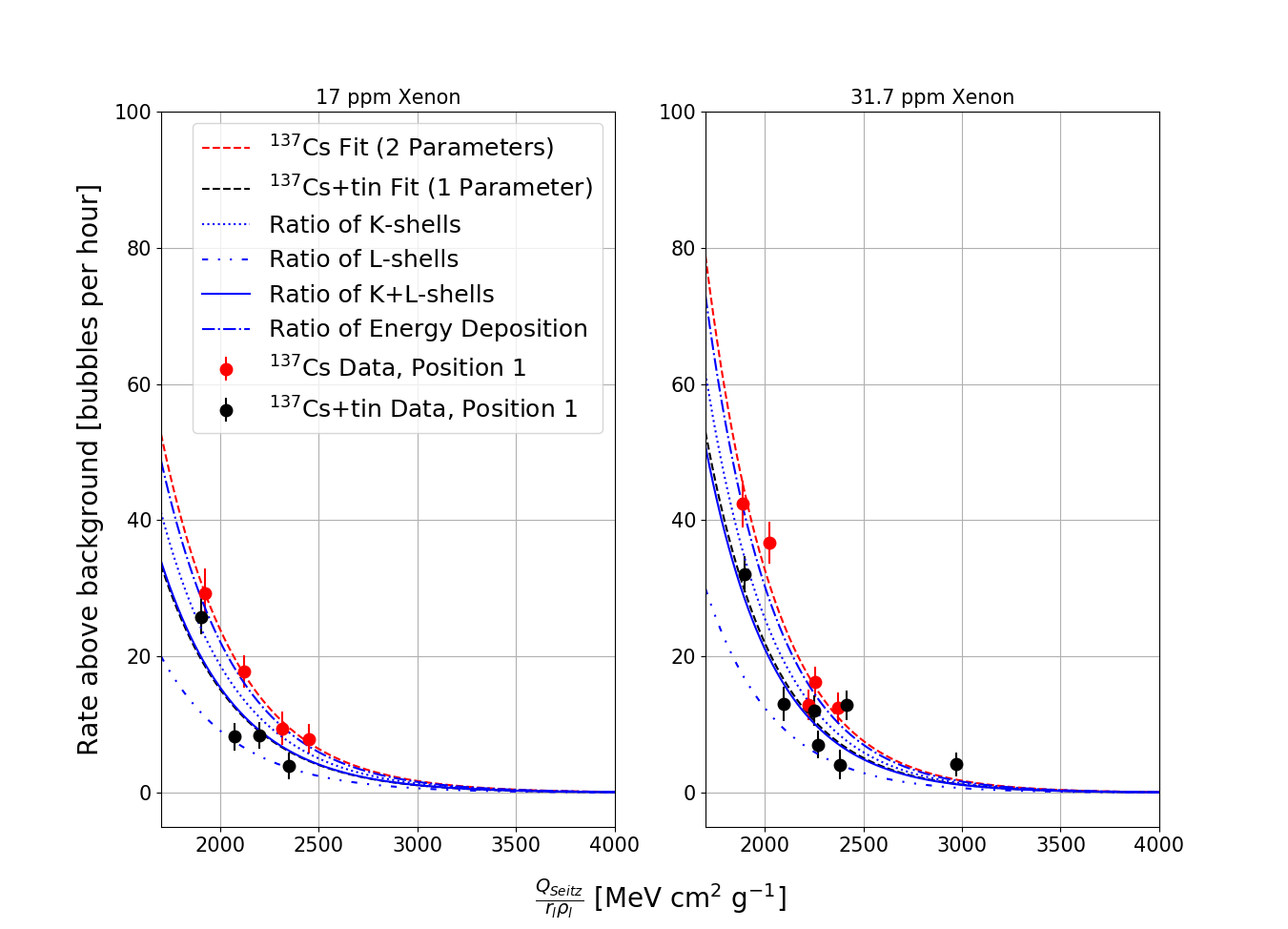}
    \caption{Bubble rates with the \cs\ source in position 1, with and without tin. Compare to figure \ref{fig:ratioseu}; with the \cs\ source, the differences between the ratios of the expected rates for vacancies in each inner shell is more apparent, but the ratios are much closer to one, so the tin does not have a large effect on the rate overall. In these data sets, the hypothesis that both K- and L-shell vacancies contribute to bubble nucleation is slightly preferred, however the uncertainties on the ratios overlap significantly so no firm conclusion is made.}
    \label{fig:ratioscs}
\end{figure}

Attempts to verify which orbital's vacancies contribute more to nucleation were inconclusive; for \eu, the ratio of rates with and without tin is essentially the same whether we consider only K-shells, only L-shells, or both K- and L-shell events, and though the ratios are slightly more distinguishable for \cs, the ratios are closer to 1 so the overall rates of bubbles with and without tin are less separated in general and the uncertainties on the ratios are significant. For three of the four \cs\ data sets (17 and 31.7~ppm for each of positions 1 and 2), the $\chi^2$ statistic slightly prefers the sum of K- and L-shell vacancies as the better fit, in the fourth case the K-shell-only fit is slightly preferred. Figure \ref{fig:ratioscs} shows the rates with \cs\ in Position 1 for 17 and 31.7~ppm xenon, similar to figure \ref{fig:ratioseu}.

\subsection{Importance for Future Experiments}
Using these results, which clearly indicate that high-Z contamination increases a \cf\ bubble chamber's sensitivity to background gammas, we can set a limit on the allowed high-Z contamination in future experiments. For example, the PICO-500 bubble chamber will contain about 250~L (350~kg) of \cf\ \cite{pico500size}, and operate at similar thresholds to those tested in the DBC. Background gamma interactions in the \cf\ from the SNOLAB environment are expected at a rate of roughly 5~Hz \cite{ERpaper}. From our simulations of the DBC, roughly 1\% of gamma interactions in \cf+1~ppk~Xe are on xenon. If we assume that this value scales linearly with concentration (i.e. the fraction of gammas interacting with xenon rather than \cf\ is 10 times the concentration of xenon by mass), conservatively assume that every gamma interaction with xenon creates an inner-shell vacancy, and take our best fit value for the probability of bubble nucleation per inner-shell vacancy at the temperature and pressure of PICO-500, that experiment is required to have less than 2~ppb high-Z contamination to keep the rate of bubbles caused by gamma interactions with high-Z contaminants less than 0.1~bubbles/year. Residual xenon naturally occurring in the air is not a concern at this level; even if no xenon is removed from the 250~L of air as the chamber is pumped out, the concentration after filling with \cf\ would be about 0.06~ppb, assuming a concentration of 0.05~ppm xenon in dry air \cite{xeinair}, which results in less than 0.01~bubbles/year. Other high-Z elements are unlikely to be present in air at substantial enough concentrations to be of concern, but could be introduced into the detector via inadequately cleaned components.

\subsection{Experiments with Argon}\label{section:argonspike}
Prior to the tests with xenon, our gas fill system and noble element spike procedures were tested with research-grade argon ($Z=18$). While we do not expect comparable Auger cascade induced nucleation between argon and xenon or iodine, argon makes up a relatively significant fraction of air compared to other heavier elements, so it is the heaviest element that could conceivably be left in a detector in a substantial quantity after evacuating with a vacuum pump. If the Auger cascade nucleation effect is significant for argon, the minimum vacuum and air leak specification for future bubble chamber dark matter searches would have to be low enough to eliminate the possible gamma background bubble rate due to this effect. Thus we report the results from these exploratory argon experiments to inform the requirements of future experiments.

A total of 7 spikes of argon into the \cf\ were performed, reaching a maximum concentration of $517\pm52$~ppm. Argon's K-Shell binding energy is 3.2~keV, about an order-of-magnitude lower than xenon and iodine, and much closer to our Seitz threshold for the test. Limited data sets were taken, using only the \cs\ source in position 1, at 20\degc\ and pressure setpoints of 35 and 40~psia, nominal Seitz thresholds 1.58 and 1.88~keV, respectively. Since PT4 was unreliable for the argon data, we must use PT5 for the pressure in the Seitz threshold calculation, and use expanded error bars due to the pressure differential usually seen between the hydraulic volume and the active fluid. Despite the limitations of this data, we can look for evidence of Auger cascade nucleation or other increased ER sensitivity in this range of argon contamination.

\begin{figure}
    \centering
    \includegraphics[width=\textwidth]{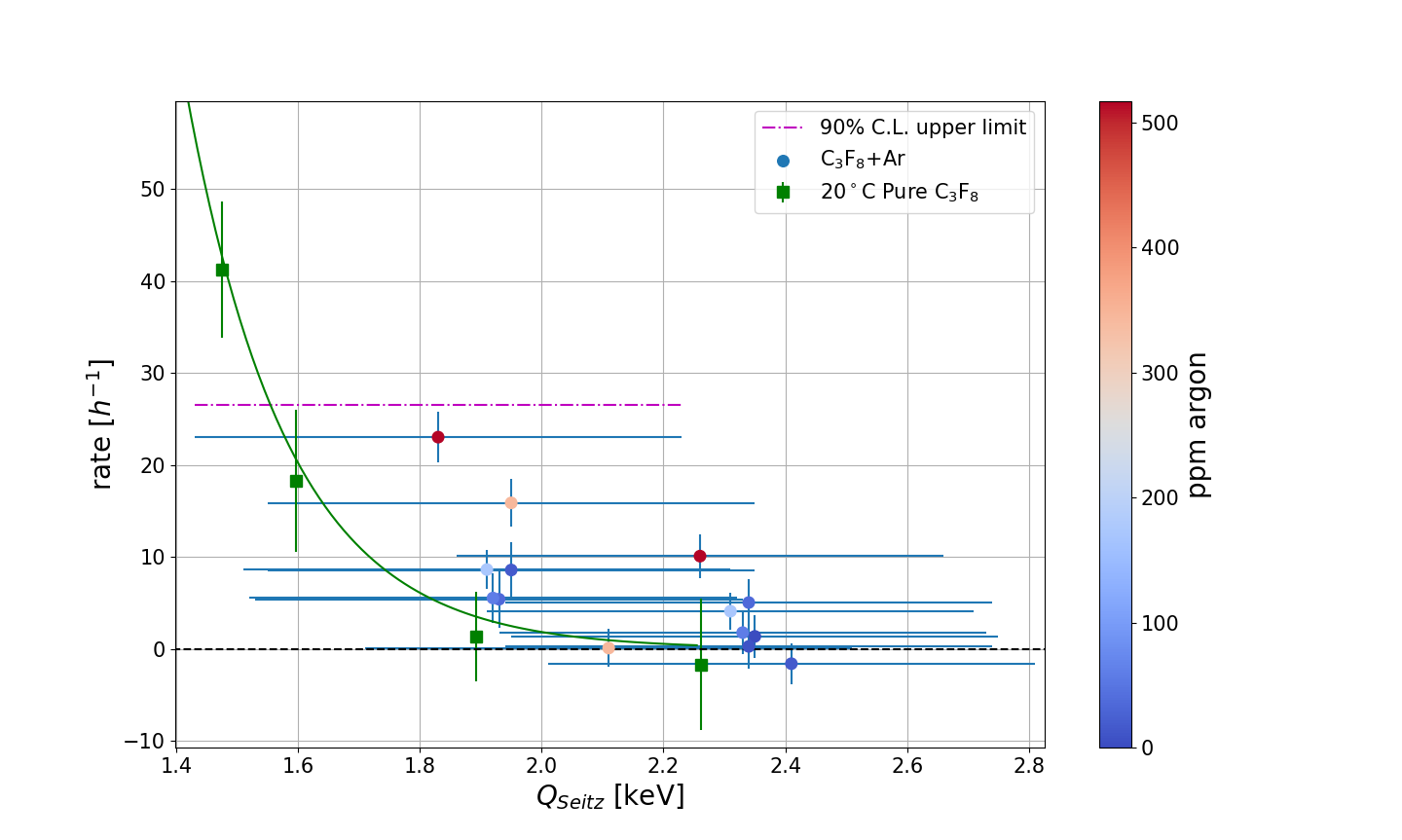}
    \caption{The DBC bubble rates when exposed to the \cs\ source, with argon contamination in the \cf\ active fluid compared to pure \cf\ at similar settings from ref. \cite{DBCpaper} in green. The data with argon are consistent with previously measured gamma rates, but we conservatively set an upper limit on the rate of bubbles per argon K-Shell photoabsorption without subtracting the rate expected from \cf\ ionization, since the thresholds are not well known. The magenta dot-dashed line is the 90\% confidence level upper limit on the bubble rate, using the lowest threshold point at the highest argon concentration.}
    \label{fig:argonrates}
\end{figure}

The actual Seitz thresholds at the mean pressures and temperatures during the test were between 1.83 and 2.41~keV. Based on the estimated $\sim5$~psid differential pressure across the bellows, we show horizontal error bars $\pm0.4$~keV. In this \qseitz\ range at 20\degc, we observe some nucleation by \cf\ ionization, which obscures our search for argon Auger cascade nucleation somewhat. Figure \ref{fig:argonrates} shows the \cf+Ar data of this test, compared to the 20\degc\ \cs\ data in pure \cf\ using the same apparatus, from ref. \cite{DBCpaper}. We observe no definitive evidence of Auger cascade nucleation in this data. Conservatively, we set an upper limit on the probability of bubble nucleation per argon K-Shell vacancy without subtracting the expected rate due to \cf\ ionization, since the threshold is not precisely known, using the measured rate $23.0\pm2.7\text{~h}^{-1}$ at \qseitz$=1.83$~keV with $517\pm52$~ppm of argon. From the simulation, the rate of argon K-Shell photoabsorptions is $4.05\pm0.52\text{~h}^{-1}\text{~ppm}^{-1}$. At the 90\% C.L., this results in an upper limit of <0.014~bubbles per K-Shell event. While this limit may not seem stringent, it is worth noting that argon K-Shell vacancies remain rare, even at such high concentrations and gamma fluxes, so the effect of Auger cascade nucleation from argon is unlikely to dramatically increase ER sensitivity at concentrations we might expect from atmospheric argon leftover after pumping low-background bubble chambers' active volumes to modest vacuum before target fluid fills.

\section{Conclusions}
We have independently verified the predictions of the Auger cascade bubble nucleation hypothesis of PICO's ER reanalysis in ref. \cite{ERpaper}. As bubble chambers aim towards detection of lower energy nuclear recoil events, understanding backgrounds associated with contaminants and background radiation becomes paramount, thus, the studies in ref. \cite{ERpaper} and this work can be used to set limits on the allowable gamma flux and the amount of contaminating elements allowed in bubble chamber target fluids for low-background experiments. Our data provide a precise calibration of the response of \cf\ bubble chambers to xenon and iodine Auger cascades at Seitz thresholds between 1 and 4~keV, where future PICO dark matter bubble chambers plan to search for WIMPs. We have also demonstrated that contamination of argon does not increase the sensitivity of the DBC to gammas from \cs\ even at concentrations $\mathcal{O}(100)$~ppm. If such future bubble chambers are kept clean of high-Z elements, the pure \cf\ model will apply, but even ppm-level contamination dramatically increases the ER response at low Seitz thresholds. For this reason, it will be important to carefully study ER response of any future bubble chamber target fluids, especially those containing high-Z elements such as the future 10~kg scintillating bubble chamber being constructed by the SBC Collaboration, which will contain a target fluid made up of argon with $\sim$100~ppm xenon, and operate at sub-keV Seitz thresholds \cite{SBCcevns, SBCICHEP}.

\acknowledgments
We would like to thank the PICO collaboration, whose technical expertise in bubble chamber operation made this work possible. In particular, we thank D.~Baxter for helpful conversations and early simulation efforts while setting up the xenon spike tests. The work of M.~Bressler is supported by the Department of Energy Office of Science Graduate Instrumentation Research Award. This work was partially supported by the Department of Energy Office of Science, Office of High Energy Physics Grant No. DE-SC0017815.

\bibliography{main.bib}
\end{document}